\documentclass[12pt,twoside]{article}
\usepackage{fleqn,espcrc1}

\usepackage{graphicx}
\usepackage[figuresright]{rotating}

\newcommand{\AmS}{{\protect\the\textfont2
  A\kern-.1667em\lower.5ex\hbox{M}\kern-.125emS}}

\title{\bf Nucleonic resonance excitations and ``missing resonances"
        in the $\omega$ meson photoproduction}

\author{Q. Zhao\address{\it Institut de Physique Nucl\'eaire, 
        F-91406 Orsay Cedex, France}
        \thanks{e-mail address: zhaoq@ipno.in2p3.fr} }

\begin{document}

\maketitle

\begin{abstract}
In this work, an improved quark model approach to the $\omega$ meson photoproduction 
with an effective Lagrangian is reported. The {\it t}-channel 
{\it natural} parity 
 exchange is consistently taken into account through the Pomeron exchange,  
while the {\it unnatural} parity exchange is described by the $\pi^0$ 
exchange. It shows that 
with very limited number of parameters, all the available experimental data
in the low energy regime can be consistently accounted for in this framework. 
Effects from the intermediate nucleonic resonances are investigated
in several polarization observables. 
The sensitivities of these observables to the resonance effects 
are probably the best way for identifying those
so-called ``missing resonances" in the baryon spectrum.
\end{abstract}

\section{Introduction and the model}
The non-relativistic constituent quark model (NRCQM) predicts a much richer 
baryon spectrum than that observed in $\pi N\to \pi N$ scatterings. 
This arises the question of the existences of the 
``missing resonances". Alternatively, it is argued that the ``absence"
of those resonances might originate from their weak 
couplings to the $\pi N$ channel. 
Therefore, efforts are also made to search for signals 
of those ``missing resonances" in other decay channels, 
for instance, $\gamma N$, $\eta N$, $K\Sigma$, $K\Lambda$, $\omega N$, 
$\rho N$, etc. This revives the study of vector meson 
($\omega$, $\rho^0$, $\rho^\pm$, $\phi$ and $K^*$)
photo- and electroproduction
near threshold in both experiment and theory.

The main feature in vector meson photoproduction is the dominant 
diffractive phenomenon at small momentum transfer.  
In the $\omega$ meson photoproduction, it has been shown in
the experimental measurement~\cite{SLAC} and vector-meson-dominant model 
(VDM) calculation~\cite{FrimanSoyeur}, that the {\it unnatural} parity 
exchange contribution plays a dominant role over the {\it natural} parity
exchange from threshold $E_\gamma=1.12$ GeV to $E_\gamma\approx 3.$ GeV.
Meanwhile, deviations from the pure 
diffractive phenomenon are expected since contributions from the 
non-diffractive intermediate resonance excitations become important. 

In this work, a $\pi^0$ exchange mechanism 
is introduced to account for 
the {\it unnatural} parity exchange, 
while a Pomeron exchange model 
by Donnachie and Landshoff~\cite{Donnachie}
is included to describe the {\it natural} parity exchange. 
The commonly used couplings, $g^2_{\pi NN}/4\pi=14$ 
and $g^2_{\omega\pi\gamma}=3.315$
 are employed for the $\pi NN$ and $\omega\pi\gamma$ vertices, respectively, 
 while an exponential form factor $e^{-({\bf q}-{\bf k})^2/6\alpha^2_\pi}$
comes from the integrals over the 
spatial wavefunctions in the harmonic oscillator basis
for these two vertices. 
${\bf k}$ and ${\bf q}$ are the momenta of the incoming photon and outgoing 
$\omega$ meson, respectively, in the c.m. system.
$\alpha_\pi=200$ MeV is the only parameter introduced for the $\pi^0$ exchange
terms. In the {\it natural} parity exchange model, the Pomeron 
mediates the long range interaction between a confined quark 
and a nucleon, and behaves rather like a $C=+1$ isoscalar photon. 
The only parameter $\beta_0=1.27$, 
which describes the coupling strength of the Pomeron 
to the constituent quark, is determined by experimental data at high energies. 
In this way, we believe that the {\it natural} and {\it unnatural} parity
exchanges have been reasonably estimated. 
We emphasize that a reliable description of the non-resonance contributions 
at forward angles is the prerequisite for studying roles played 
by the {\it s}- and {\it u}-channel resonances. 

For the {\it s}- and {\it u}-channel resonance contributions, 
 an effective Lagrangian is introduced into
 the quark-vector-meson coupling in the quark model~\cite{plb98,prc98}: 
\begin{equation} \label{Lagrangian}  
L_{eff}=-\overline{\psi}\gamma_\mu p^\mu\psi+\overline{\psi}\gamma_
\mu e_qA^\mu\psi +\overline{\psi}(a\gamma_\mu +  
\frac{ib\sigma_{\mu\nu}q^\nu}{2m_q}) \phi^\mu_m \psi,
\end{equation}  
where $\psi$ and $\overline{\psi}$ represent the quark and antiquark
fields, respectively, and $\phi^\mu_m $ denotes the vector meson 
field. The two parameters, 
$a$ and $b$ represent the vector and tensor couplings 
of the quark to the vector meson, respectively,
and $m_q=330$ MeV is the constituent quark mass. 

In the $SU(6)\otimes O(3)$ symmetry limit, 
those low-lying contributing states with $n\leq 2$ in the harmonic oscillator 
basis are listed in Table~\ref{table-1}, while those excited states 
with $n>2$ are safely treated degenerate with $n$. 
It should be noted that the Moorhouse
selection rule~\cite{moor} has eliminated 
those states belonging to NRCQM representation
${\bf [70, ^4 8]}$ from contributing. 

\begin{table}[htb]
\caption{ Resonances contributing in the {\it s}-channel, with their 
assignments in the $SU(6)\otimes O(3)$ symmetry limit. 
$M_R$  and $\Gamma_T$ are the masses and total widths of resonances
which are taken from Ref.~\cite{PDG}.
} 
\protect\label{table-1}
\begin{center}
\begin{tabular}{cccc}
\hline
Resonances & $SU(6)\otimes O(3)$ & $M_R$ (MeV) & $\Gamma_T$ (MeV) \\
\hline
$S_{11}(1535)$ & $N(^2P_M)_{{\frac 12}^-}$ & 1535 & 150\\
$D_{13}(1520)$ & $N(^2P_M)_{{\frac 32}^-}$ & 1520 & 120\\
$P_{13}(1720)$ & $N(^2D_S)_{{\frac 32}^+}$ & 1720 & 150\\
$F_{15}(1680)$ & $N(^2D_S)_{{\frac 52}^+}$ & 1680 & 130\\
$P_{11}(1440)$ & $N(^2S^\prime_S)_{{\frac 12}^+}$ & 1440 & 350\\
$P_{11}(1710)$ & $N(^2S_M)_{{\frac 12}^+}$ & 1710 & 100\\
$P_{13}(1900)$ & $N(^2D_M)_{{\frac 32}^+}$ & 1900 & 250\\
$F_{15}(2000)$ & $N(^2D_M)_{{\frac 52}^+}$ & 2000 & 250\\
\hline
\end{tabular}
\end{center}
\end{table}

The constraint on the two parameters, $a$ and $b$, 
for the resonance contributions
comes from the large angle behavior in the differential cross sections. 
At large angles, it is the {\it s}- and {\it u}-channel resonance 
contributions that play dominant roles over the {\it t}-channel 
 contributions. To reproduce the bump structures 
observed in experiments~\cite{ABBHHM,SLAC,SAPHIR}, we find that
$a=-0.8$ and $b=-1.6$ give an overall fitting for all the data 
available~\cite{zhao}. 

\section{Observables}
\begin{figure}[htb]
\begin{minipage}[t]{80mm}
\includegraphics[width=20pc]{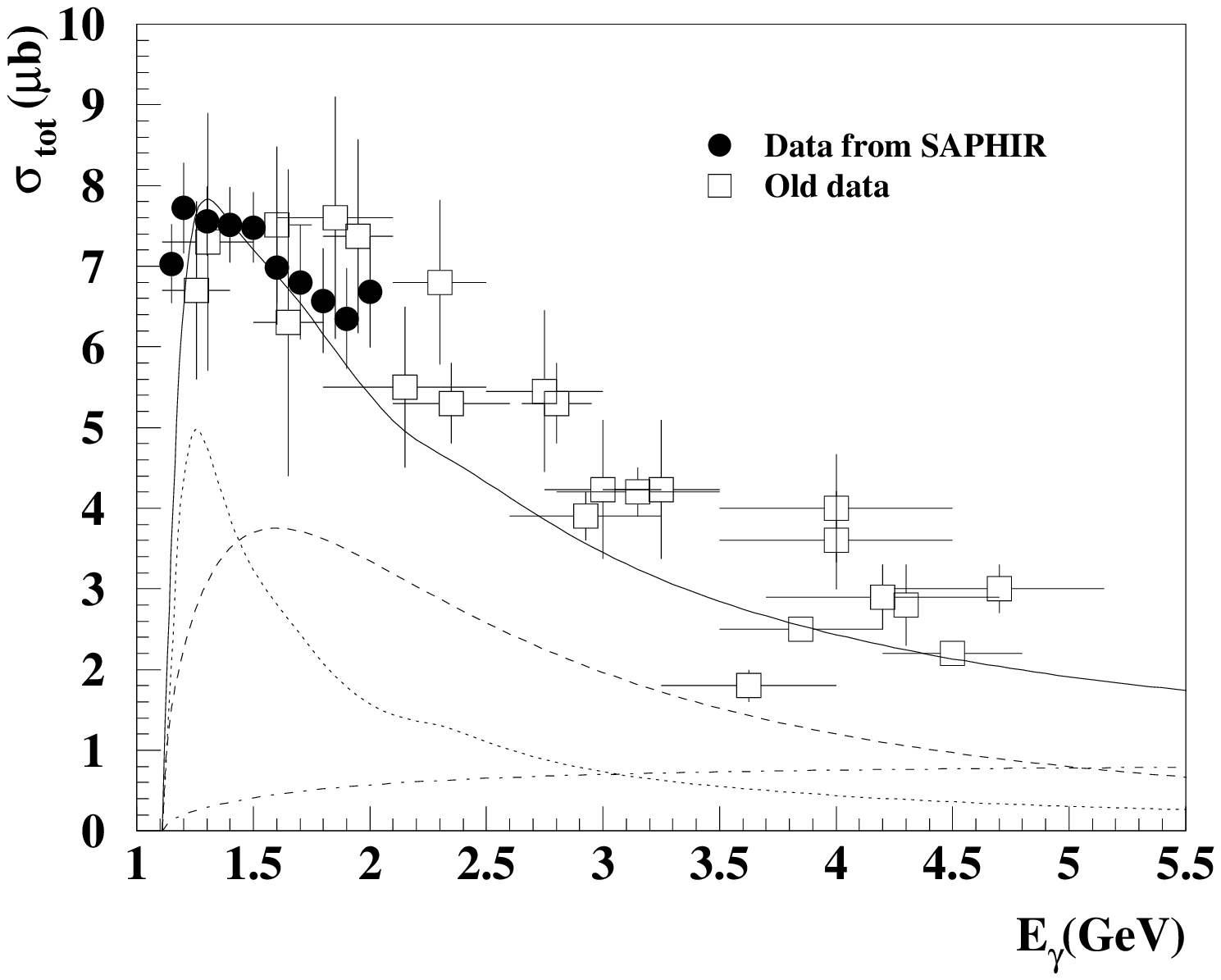}
\caption{Total cross section. Notations are given in the text.}
\label{fig:total-x}
\end{minipage}
\hspace{\fill}
\begin{minipage}[t]{70mm}
\includegraphics[width=16pc]{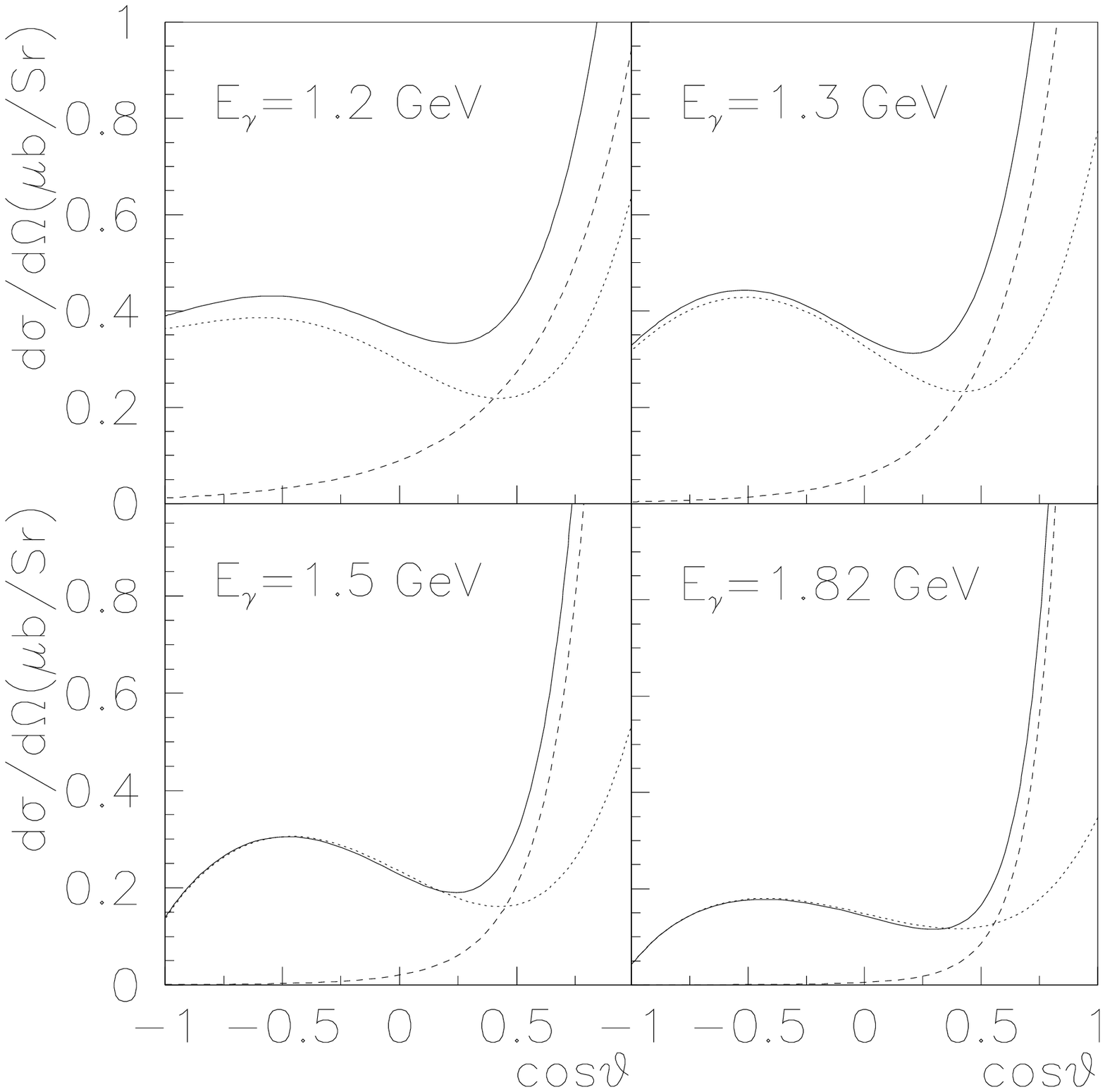}
\caption{ Angular distributions predicted by this model.}
\label{fig:distribution}
\end{minipage}
\end{figure}

In Fig.~\ref{fig:total-x}, the total cross section (full curve)
calculated in this model is presented. 
The {\it unnatural} parity exchange contribution is accounted for 
by the $\pi^0$ exchange (dashed curve) over a large energy regime.
 With the energy increasing, the Pomeron exchange (dot-dashed), which 
accounts for the {\it natural} parity exchange contributions 
becomes more and more important, and finally dominates the cross sections 
at high energies. 
The inclusion of the Pomeron exchange terms improves 
the previous calculations~\cite{FrimanSoyeur,prc98}
significantly when the photon energies are above the resonance region 
(from threshold to about 2.2 GeV). 
The most interesting feature from the resonance 
contributions (dotted curve) is that, although the $\pi^0$ exchange
plays a dominant role over a large energy regime, it is the resonance 
contributions
that definitely dominate over the other two processes 
from threshold to about $E_\gamma=1.5$ GeV. This feature accounts 
for the flattened angular distributions 
near threshold observed in CLAS experiment~\cite{CLAS}. 

In Fig.~\ref{fig:distribution}, the angular distributions for different energy scales 
are presented. 
Explicitly, it shows that the {\it s}- and {\it u}-channel resonance
contributions (dotted curves) produce the bump structures at large angles, 
while the $\pi^0$ plus Pomeron exchange (dashed curves) account for
the forward angle diffractive behaviors.
The energy evolution of the large angle behavior is quite 
sensitive to the relative sign and strength between $a$ and $b$, 
which makes the model prediction absolutely non-trivial. 
We find that
that the predicted angular distributions are in good agreement 
with the preliminary CLAS data~\cite{CLAS}.

\vspace*{-1.cm}

\begin{figure}[htb]
\begin{minipage}[t]{70mm}
\includegraphics[width=20pc]{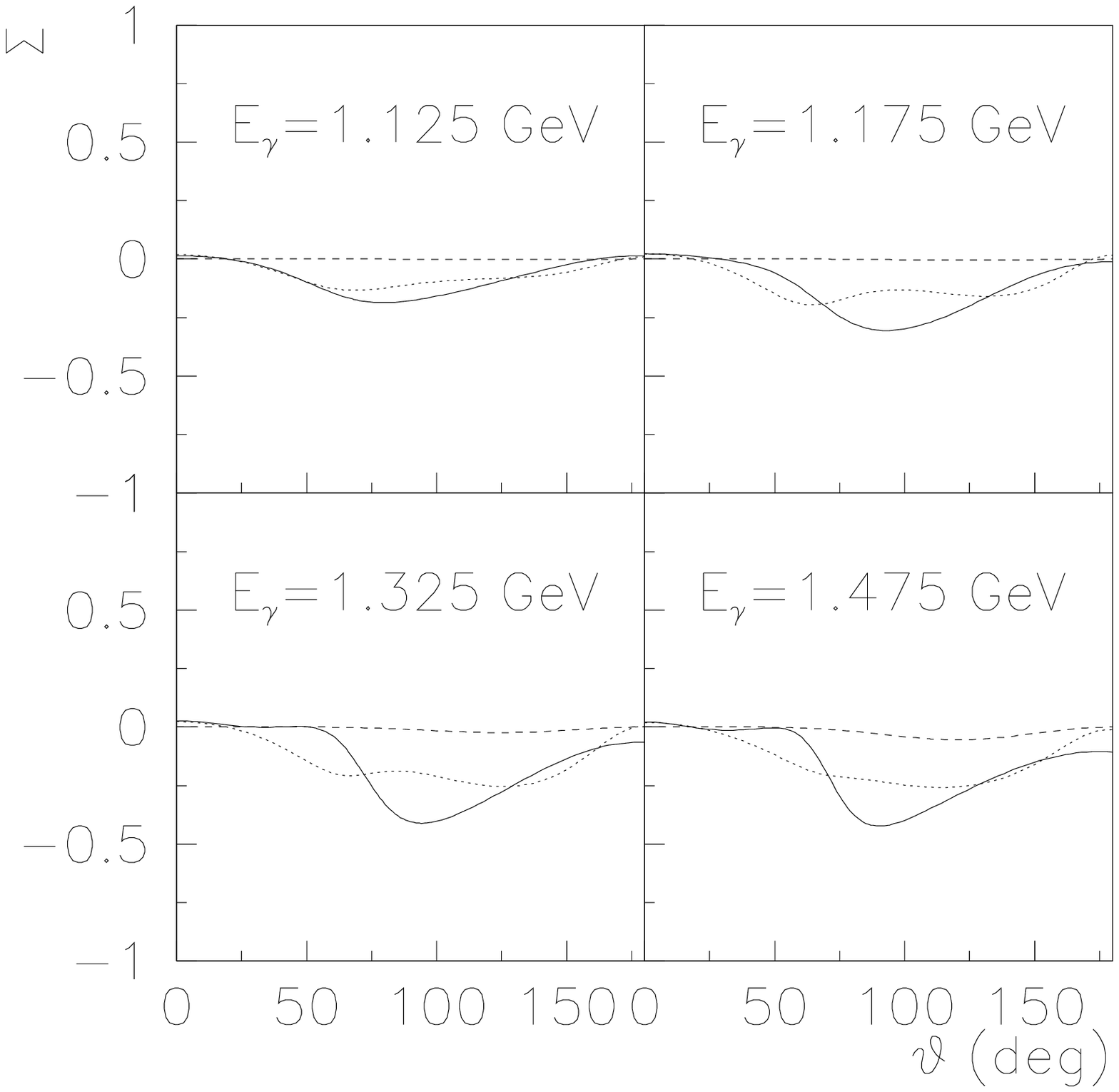}
\caption{Beam polarization asymmetry within GRAAL energy scope. 
Notations are given in the text.}
\label{fig:beam}
\end{minipage}
\hspace{\fill}
\begin{minipage}[t]{80mm}
\includegraphics[width=16pc]{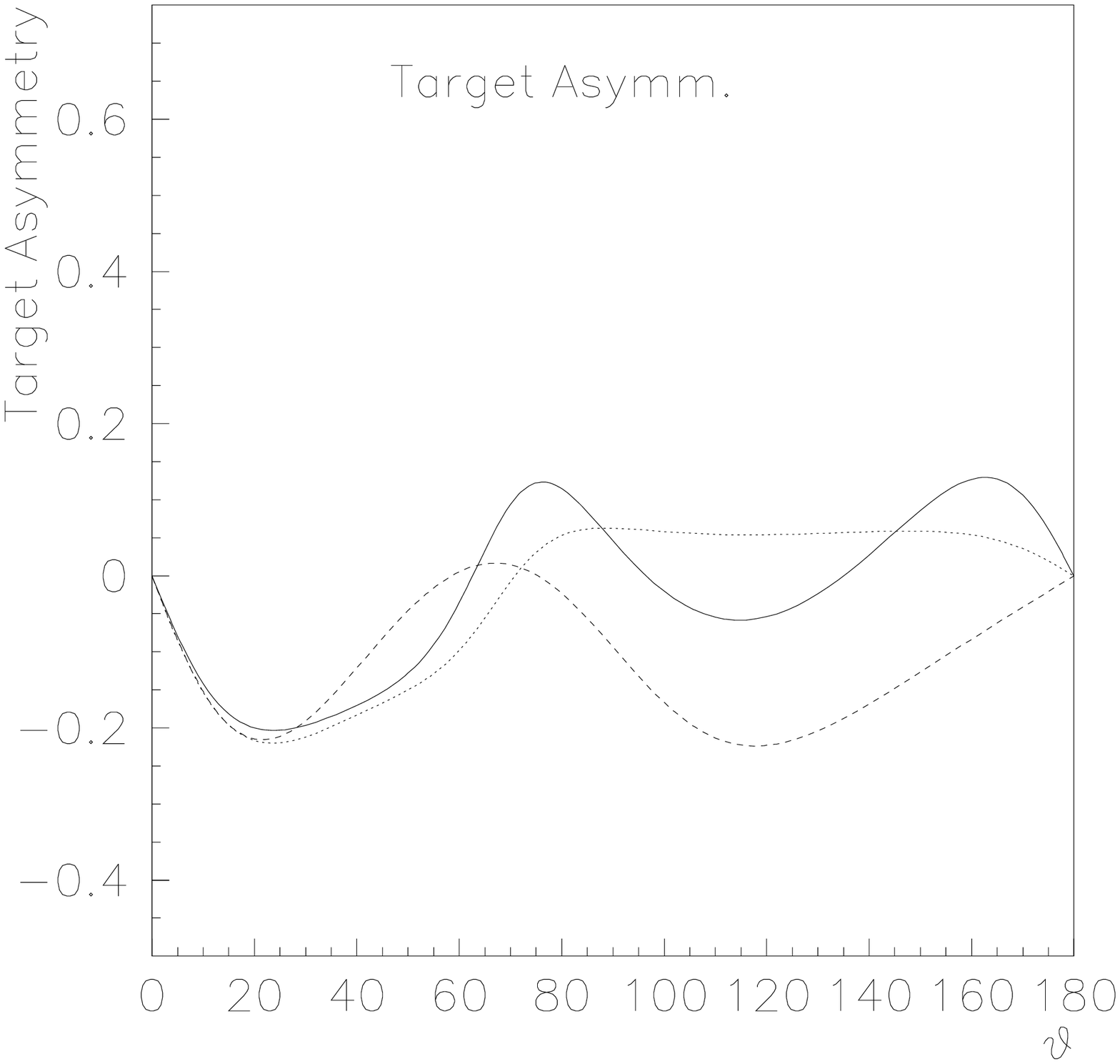}
\caption{Target polarization asymmetry predicted at $E_\gamma=1.6$ GeV. 
Notations are given in the text.}
\label{fig:target}
\end{minipage}
\end{figure}

Our predictions for the beam polarization 
asymmetries are presented in Fig.~\ref{fig:beam}. The dashed curves
represent the asymmetries with only the {\it t}-channel exchange terms, 
namely, the $\pi^0$ and Pomeron exchanges. It can be justified that 
when only pure {\it unnatural} or {\it natural} parity 
exchanges contribute to the amplitudes, the beam polarization 
asymmetries will be zero. This rule also holds 
when only the $\pi^0$ 
and Pomeron exchanges contribute in $\gamma p\to \omega p$.
Since here the $\pi^0$ 
exchange and Pomeron exchange contribute to the real and imaginary 
amplitudes, respectively, 
no interfering terms (real-imaginary components) appear
in the beam polarization asymmetries (the beam asymmetry only involves 
the real-real or imaginary-imaginary interfering components). 
This feature makes the beam polarization asymmetry interesting 
since the  {\it s}- and {\it u}-channel resonance interferences 
can be reflected by the non-zero asymmetries. 
With the {\it s}- and {\it u}-channel resonance contributions included, 
the beam polarization asymmetries are shown by the full curves. 
We find that a clear nodal structure 
appears at about $90^\circ$. Also, 
a flattened behavior with very small asymmetries
is also found in the forward angles between $0^\circ$ and $60^\circ$.
Similar feature has been observed in the preliminary data from GRAAL.
The energy evolution of the asymmetries will provide a challenge 
for a model prediction. 

In the $SU(6)\otimes O(3)$ symmetry 
limit, it shows that two resonances 
$P_{13}(1720)$ and $F_{15}(1680)$,
which are assigned to representation ${\bf [56, ^2 8]}$
with $n=2$, play dominant roles over contributions from other states. 
The dotted curves in Fig.~\ref{fig:beam} show the effects without contributions
from $P_{13}(1720)$. This dominating behavior can be accounted for 
by the non-vanishing longitudinal, and electro-like transitions 
at the meson coupling vertex. In comparison with  
$P_{13}(1720)$ and $F_{15}(1680)$, the contributions from 
$P_{13}(1900)$ and $F_{15}(2000)$, which belong to representation
${\bf [70, ^2 8]}$ with $n=2$,
turn out to be quite small. In the quark model symmetry 
limit, only the magnetic-like transition is permitted for 
these two states. 

\vspace*{-1.5cm}

In Fig.~\ref{fig:target}, the target 
polarization asymmetries (full curve) are shown at $E_\gamma=1.6$ GeV. 
The dashed curves represent the asymmetries from the {\it t}-channel 
exchanges, i.e. the $\pi^0$ plus Pomeron exchange. In comparison with 
the full curves in which the resonance contributions 
are taken into account, we find that the resonance contributions 
play significant roles at large angles. 
Interestingly, with the energy increasing, the interference 
between the $\pi^0$ and Pomeron exchange  
determines the asymmetries at small angles while at large angles
it is the resonance interference that plays a dominant role. 
The dotted curve shows the effects 
when the two ``missing resonances" are eliminated from contributing.
It shows that at $E_\gamma=1.6$ GeV, the nodal structures 
at intermediate and large angles come from the interferences of the 
two ``missing resonances". The target polarization asymmetry 
measurement might be able to provide signals for the existence of 
the two ``missing resonances".

\section{Conclusion}
In summary, we report an improved quark model approach 
to the $\omega$ meson photoproduction.
With very limited number of 
parameters, all the available data 
in the low-energy region can be consistently reproduced. 
Such a framework shows great potential in the study 
of the resonance effects in vector meson photoproduction.
More detailed results will be reported elsewhere~\cite{zhao}.

\section*{Acknowledgement}
The author 
appreciates useful discussions with B. Saghai,
J.-P. Didelez, M. Guidal and P. Cole.
Warm invitation from IHEP and useful discussions with 
B.S. Zou are thanked.

\end{document}